\documentclass[11pt]{article}
\usepackage{graphicx}
\DeclareGraphicsExtensions{.jpg}

\begin{document}
%%%%%%%%%%%%%%%%%%%%
\title{{\bf{\Large Corrections to Unruh effect in tunneling formalism and mapping with Hawking effect}}}
%%%%%%%%%%%%%%%%%%%%
\author{
{\bf {\normalsize Rabin Banerjee}$
$\thanks{E-mail: rabin@bose.res.in}},\,
 {\bf {\normalsize Bibhas Ranjan Majhi}$
$\thanks{E-mail: bibhas@bose.res.in}},\, 
{\bf {\normalsize Debraj Roy}$
$\thanks{E-mail: debraj@bose.res.in}}\\
 {\normalsize S.~N.~Bose National Centre for Basic Sciences,}
\\{\normalsize JD Block, Sector III, Salt Lake, Kolkata-700098, India}
\\[0.3cm]
}
%\date{}

\maketitle
\begin{abstract}

We consider coordinate systems adapted to accelerated observers, construct a generalised Rindler metric and then adopt a specific form of it to compute the semiclassical Unruh temperature, using a WKB approximation within a Hamilton Jacobi analysis in the tunneling picture. Corrections to this temperature are next computed by going beyond the usual WKB approximation. A connection of the corrected Unruh temperature with the Hawking temperature is established. This connection is also explained through the method of Global Embedding Minkowski Spacetime (GEMS).

\end{abstract} 

\section{Introduction}

The difficulties in the definition of a general notion of vacuum were demonstrated in W. G. Unruh's 1976 paper \cite{Unruh} where, among other discussions, he showed how an accelerated observer in flat Minkowski spacetime observing the Minkowsi vacuum, will actually find a thermal spectrum. Thus the appropriate vacuum state corresponding to him would be distinct from the Minkowski vacuum. The connection between the above mentioned `Unruh effect' and the prediction by Hawking that Black Holes should radiate -- `Hawking Effect' \cite{Hawking1,Hawking2} -- was also investigated by Unruh in \cite{Unruh}.

Though \textit{a priori} it is not evident how an observation made in flat spacetime is related to an effect that occurs in the curved black hole spacetime, the key lies in the principle of equivalence. A freely falling observer near the horizon would observe locally around him a flat Minkowski metric. Also he would see a static observer in his vicinity as an observer accelerating with a constant acceleration, say $\alpha$. Now if the static observer sees the vacuum state of the freely falling observer, he would see a thermal flux at an Unruh Temperature $\frac{\hbar \alpha}{2\pi}$. Again a redshifted version of this same temperature will be seen at infinity as the Hawking Temperature $\frac{\hbar\kappa}{2\pi}$ where $\kappa$ is the surface gravity of the black hole.

An alternative way to explore the mapping of Unruh and Hawking thermal properties is to recall that any d-dimensional geometry has a higher dimensional Global Embedding Minkowski Spacetime (GEMS) \cite{Goenner}. Several examples of this mapping have been illustrated in \cite{Rosen}. This property relates the results obtained by (Hawking) detectors in curved spaces with those of (Unruh) detectors in flat spaces \cite{Deser}.

Results for the Unruh and Hawking temperatures were at first all derived from classical or semi-classical considerations. But corrections to Hawking temperature due to quantum fluctuation of spacetime near the event horizon, thereby accounting for the back reaction effect, have also been widely studied \cite{Fursaev, Page, York, Lousto, Majhi, Majhibeyond, Majhiback, Majhifermi}. This raises the interesting question as to whether the Unruh temperature would also have appropriate corrections. But this issue and its connection, if any, with the corrected Hawking temperature has not been adequately addressed in the literature \cite{Reznik}. Here in this work we address that issue following both approaches mentioned above.

The original analysis of Hawking \cite{Hawking1,Hawking2} was fairly involved. The result being remarkable, several other derivations were attempted subsequently \cite{Hartle, GibbHawk}.  But the approach that truly matches an intuitive picture as a source of radiation tunneling across the horizon in a classically forbidden process, is the tunneling formalism developed in \cite{Paddy, Wilczek}, for scalar particles in a Schwarzschild spacetime. The formalism since then has both been refined and applied to Dirac particles as well as various other black hole spacetimes \cite{Majhi, Majhifermi, Paddy2, Vanzo, Kim, Jiang, Modak, Chen, Akhmedov, Bibhas, RB}. Thus to compute corrections to the Unruh effect, we readily employ the tunneling formalism keeping in mind its intuitive clarity.

Now for the analysis of Unruh effect \cite{Unruh, Crispino}, an appropriate coordinate system for the accelerated observer is required. This is the Rindler coordinate system which covers the whole Minkowski plane in four separate coordinate patches ({\it see fig. \ref{wedgedia} \& \ref{penrose}}). Here we construct a generalised Rindler metric which can accommodate various parametrisations based on the trajectories of the accelerated observer. It can thus give any of the various forms of the Rindler metric seen in the literature \cite{Unruh, Rindler, Carrol}, through a proper choice of a single function.

We clarify several aspects of the Unruh effect, especially in the tunneling mechanism, and present new results for corrections to the Unruh effect. A mapping of the corrected Unruh and Hawking temperatures is established. We show that tunneling in Rindler coordinates occurs from region $\textrm{II}$ (the black hole like region) to $\textrm{I}$ (the `physical' region) across the Rindler horizon ({\it see fig. \ref{wedgedia}}). Later, when the near horizon form of any generic black hole metric is deduced to be a Rindler metric, we actually see that the Rindler horizon maps to the black hole horizon and we beautifully get back the picture of particles tunneling out from the inside of a black hole in a classically forbidden process. This gives a nice check, supporting our tunneling picture in the Unruh phenomenon.

Tunneling is usually employed through two different methods - the radial null geodesic method \cite{Wilczek} and the Hamilton Jacobi method \cite{Paddy, Paddy2}. Here we adopt the latter to calculate both the semi-classical Unruh temperature and corrections to it due to quantum effects. Considering scalar particle tunneling, we write the Klein Gordon equation in the Rindler metric and make the standard WKB ansatz for the wave function $\phi = e^{-i/\hbar~S(x,t)}$. In the full analysis, the single particle action $S(x,t)$ is expanded in powers of $\hbar$ as $S=S_0 + \sum_i\hbar^iS_i$. However at the first order semiclassical level, we take $\hbar \rightarrow 0$ limit and find the ingoing and outgoing modes of $\phi$. Imposing classical condition of unitary ingoing probability, we get the imaginary temporal contribution to the action and use it to compute the outgoing probability. Now using the principle of detailed balance \cite{Paddy2,Bibhas} for an arbitrary observer, we again get the outgoing probability in terms of the characteristic temperature - the Unruh temperature - and get our result by comparison. This semiclassical Unruh temperature matches the standard form of acceleration upon $2\pi$. In addition, our result also naturally incorporates the coordinate dependence in the acceleration, as different observers in the Rindler metric have different accelerations based on their coordinate positions. Now near the black hole horizon, we use our result to get a Unruh temperature for the relevant Rindler metric and redshift it to infinity to get the Hawking temperature. Exactly at the horizon, the Unruh temperature obtained is infinitely high which is expected since any finite energy process at horizon is not observable outside.

When we go beyond the first order level and keep all corrections in different powers of $\hbar$, we explicitly demonstrate by induction that all higher orders $S_i(x,t)$ in the single particle action are proportional to the first-order action $S_0$. This explicit proof was lacking previously in the literature where the general relation was only deduced from a few initial order results \cite{Majhibeyond, Majhiback, Majhifermi, Modak}. Thus now the action is just modified by a pre-factor and the whole analysis repeats the semi-classical case leading finally to the corrected Unruh temperature. Proportionality constants of the correction terms are next fixed by using this result near a Schwarzschild horizon and redshifting the corrected Unruh temperature thus found to infinity, giving the known standard form of corrected Hawking temperature. The first order proportionality constant, for the case of a Schwarzschild black hole, is now found by comparison. An analysis through GEMS also conforms to the above results.

Now we briefly explain the organization of our paper. In Section 2 we set up a generalised Rindler metric and make a specific choice of metric which is to be used in the subsequent analysis. Next, the Unruh effect is taken up in Section 3 where we calculate the semicalssical Unruh temperature in the tunneling approach of \cite{Paddy, Paddy2}, through a Hamilton-Jacobi  formalism. The physical picture of tunneling across the Rindler horizon and some subtleties are also discussed in this section. Then in Section 4 we introduce effects of quantum corrections and use the Hamilton-Jacobi approach again, but now going beyond the usual semi-classical approximation \cite{Majhibeyond, Majhiback, Majhifermi, Modak} to compute the corrected Unruh temperature to all orders in $\hbar$. In Section 5, connection between the Unruh and Hawking temperatures is discussed through two different methods (near horizon approximation and GEMS). The connection is then used to fix the proportionality constant of the correction term, upto first order . Finally we conclude with a discussion of the results in Section 6. We give two appendices: in A we show how some standard Rindler metrics used in literature can be obtained from the generalised Rindler metric, and in B we do the tunneling calculation in the generalised form of the Rindler metric. Two figures have been included to further explain the generalised Rindler metric which has been constructed. Fig. \ref{wedgedia} shows the Rindler wedges in the Minkowski plane and explains where the tunneling occurs. Fig. \ref{penrose} shows the conformal Carter-Penrose diagram for the first Rindler wedge of the generalised metric.

\section{Accelerated observer in Minkowski spacetime -- Rindler\\coordinates}
        
\begin{figure}[t]
\centering
\includegraphics[angle=0, height=100mm,width=126mm]{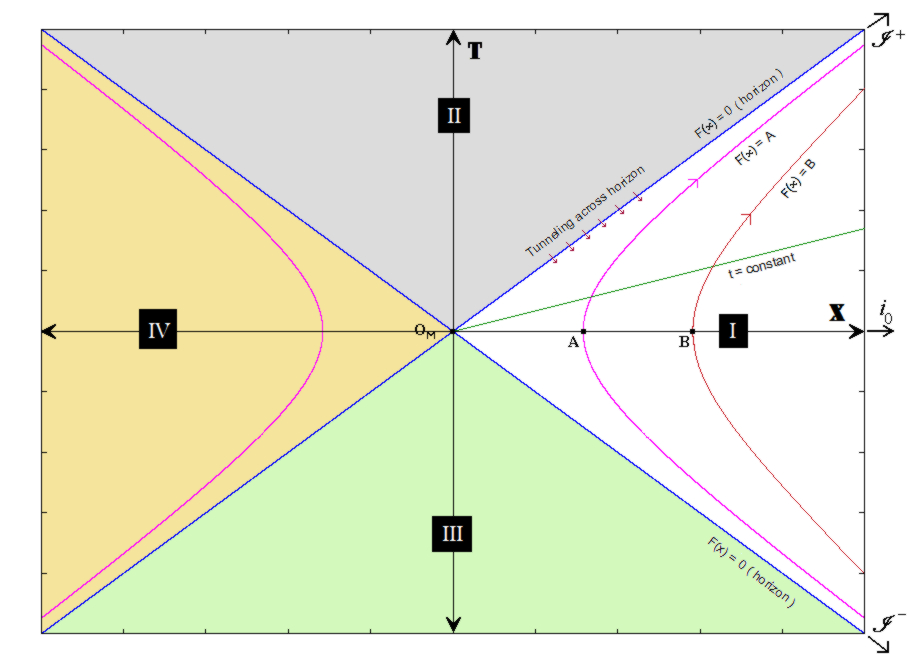}
\caption{{\it The Rindler wedges shown on the Minkowski plane. $\textrm(${\bf X},{\bf T}$\textrm)$ and ($x,t$) are Minkowski and Rindler coordinates. The infinities lie outside the diagram, towards the directions shown. Tunneling occurs from region $\textrm{II}$ to region $\textrm{I}$.}}
\label{wedgedia}

\includegraphics[angle=0, height=70mm,width=80mm]{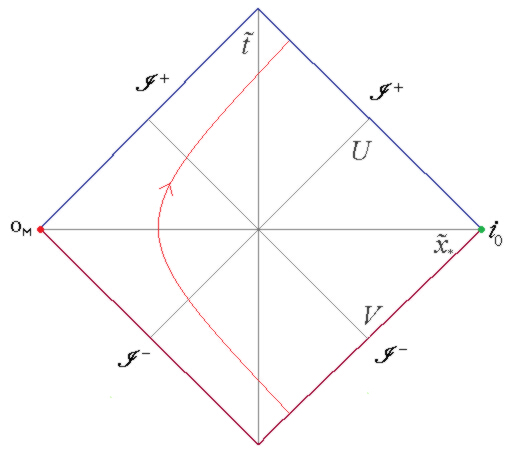}
\caption{\it Carter-Penrose diagram - Conformal representation of the Rindler wedge $\textrm{I}$ showing the position of the infinities. $\tilde t$ is the inverse tangent of rindler time $t$ and $\tilde x_*$ is the inverse tangent of a tortoise like space coordinate constructed out of the Rindler $x$. The null infinities shown are only those portions of the corresponding infinities of the Minkowski plane that lie inside the wedge $\textrm{I}$. Minkowski origin $O_M$ is seen on one vertex of the diagram. The directed geodesic shown is that of a Rindler observer.}
\label{penrose}
\end{figure}

An observer accelerating with a constant acceleration $\alpha$ along the $X$-axis in Minkowski spacetime follows a hyperbolic trajectory \cite{Mould}
\begin{eqnarray}
X^2 - T^2 = \frac{1}{\alpha ^2}
\label{hyper}
\end{eqnarray}
as predicted by special relativity, where $T$ and $X$ are the Minkowski time and space coordinates. The parametrized trajectory equation is
\begin{eqnarray}
T &=& \frac{1}{\alpha} \sinh(\alpha\tau)\nonumber\\
X &=& \frac{1}{\alpha} \cosh(\alpha\tau),
\label{accobs}
\end{eqnarray}
where $\tau$ is the proper-time of the accelerating observer. Here, depending on the sign of $\alpha$, the trajectories cover two sections of the Minkowski Spacetime ({\it see figure \ref{wedgedia}})
\begin{eqnarray}
\textrm{I.} \quad & X>0, \; \vert X\vert\geq\vert T \vert \qquad \alpha > 0\nonumber ,\\
\textrm{IV.} \quad & X<0, \; \vert X\vert\geq\vert T \vert \qquad \alpha < 0.
\label{wedges}
\end{eqnarray}
It is to be noted that, physically a positive $\alpha$ will mean acceleration for $T>0$ and a deceleration for $T<0$, with the scenario being reversed for negative $\alpha$.

An appropriate coordinate system adapted to the trajectory of this accelerated observer is the Rindler Coordinates $(t,x,y,z)$ defined through the transformation (in region $\textrm{I}$)
\begin{eqnarray}
T &=& F(x) \sinh(a t)\nonumber\\
X &=& F(x) \cosh(a t)\nonumber\\
Y &=& y; \qquad Z\;=\;z.
\label{rtrns}
\end{eqnarray}
The restrictions (\ref{wedges}) result in the coordinates $(t, x)$ thus defined to cover only a section of the Minkowski Coordinates $(T, X)$. Other sections can be covered by analogously defined separate coordinate pairs. These sections are referred to as {\it `Rindler Wedges'}. Here $F(x)$ is a monotonic, analytic, positive definite function of `$x$' which parametrizes different trajectories for different values of acceleration and `$a$' is a constant in the $(t,x,y,z)$ coordinates. The coordinates $x$ and $t$ are related to the acceleration and the proper time of the accelerated observer through
\begin{eqnarray}
\alpha &=& \frac{1}{F(x)}\nonumber\\
\tau &=& \frac{a}{\alpha}~t.
\label{accn}
\end{eqnarray}
We see from the above relations that for an observer with coordinate time equal to proper time, the acceleration $\alpha$ is equal to the parameter $a$. 

Finally, the transformed metric obtained from the Minkowski metric
\begin{equation}
ds^2 = -dT^2 + dX^2 + dY^2 +dZ^2
\label{mink}
\end{equation}
is the generalized Rindler metric
\begin{equation}
ds^2 = -a^2 [F(x)]^2 dt^2 + [F'(x)]^2 dx^2 + dy^2 + dz^2.
\label{genrind}
\end{equation}
Different forms of the Rindler metric often seen in literature ({\it see for example} \cite{Unruh, Rindler, Carrol}) can all be obtained through an appropriate choice of the function $F(x)$. Any metric then is in Rindler form only if we can get a well-behaved function $F(x)$ and a {\it constant} `$a$' such that it can be expressed in the form of (\ref{genrind}). For example, for the choice of
\begin{eqnarray}
F(x)&=&\frac{2\sqrt x}{\sqrt \Lambda}\;=\;\frac{1}{\alpha}\nonumber\\
\& \quad a &=& \frac{\Lambda}{2},
\label{ourparam}
\end{eqnarray}
where $\Lambda$ is some constant parameter, we get the specific form of the Rindler metric
\begin{eqnarray}
ds^2 = -\Lambda x dt^2 + \frac{dx^2}{\Lambda x} + dy^2 + dz^2
\label{rind1}
\end{eqnarray}
which is used in {\it Sections \ref{SecU} \& \ref{SecUCorr}} to calculate, first, the standard Unruh temperature and, next, corrections to it. Some more examples of different choices of $F(x)$ leading to different Rindler metrics have been discussed in Appendix A. 

In Figure \ref{wedgedia} we show the hyperbolic trajectories of the accelerated observer. The two hyperbolae for $F(x)=A$ and $B$ with $A<B$ correspond to two different accelerations $1/A$ and $1/B$. Thus higher values of acceleration correspond to hyperbolae which intersect the Minkowski $X$ axis at points increasingly closer to the Minkowski origin $O_M$. Observers with infinite acceleration $\alpha$, travel along the lines $F(x)=0$ (see \ref{accn}), which form the accelerated horizon of the Rindler observers. In Minkowski coordinates, the hyperbolic trajectory equation (\ref{hyper}) gives the same horizon as $X=\pm T$. We note that a naive, direct substitution of $F(x)=0$ in the transformation (\ref{rtrns}), identifies only the single point describing the Minkowski origin as the horizon. However the transformation (\ref{rtrns}) itself then becomes non-invertible and hence the conclusion would be erroneous. It can be shown \cite{Mould} that no information interchange can occur between regions $\textrm{I}$ and $\textrm{IV}$. Region $\textrm{II}$ is like a blackhole, i.e. no information from there can reach the Rindler observers in region $\textrm{I}$, while the contrary is possible. So the accelerated horizon $F(x)=0$ acts like an event horizon of a black hole. Similarly, region $\textrm{III}$ is like a white hole. The past and future null infinities as well as the spacelike infinity of the Rindler observers lie in the directions as indicated.

In the later two sections we use the specific Rindler metric (\ref{rind1}) to calculate both the semiclassical Unruh temperature and corrections to it due to quantum effects (see Appendix B for tunneling in the generalised metric (\ref{genrind}) ). Since we are interested only in the $x-t$ sector of the metric, we will suppress the $y$ and $z$ coordinates henceforth.

\section{Unruh effect by quantum tunneling}

This section is devoted to calculate the Unruh temperature by tunneling method, where quantum mechanically, a particle can tunnel across the horizon in an otherwise classically forbidden direction. Here we consider tunneling of a massless scalar particle through the horizon, from region $\textrm{II}$ to region $\textrm{I}$ as shown in Fig.\ref{wedgedia}. Since in the classical limit ($\hbar\rightarrow 0$) the ingoing probability is unity and the outgoing probability is zero, the ingoing single particle action is real while the outgoing action is complex. The tunneling rate is proportional to the exponential of the imaginary part of the outgoing action, which is compared with the Boltzmann factor through the `{\it principle of detailed balance}' \cite{Paddy2,Bibhas} leading to the Unruh temperature. Therefore the important part of this method is to calculate the imaginary part of the outgoing one particle action. There are two ways to do this: (i) Radial null geodesic method \cite{Wilczek} and (ii) Hamilton Jacobi method \cite{Majhibeyond, Paddy, Paddy2}. We will consider the Hamilton-Jacobi method, where the analysis in this section will be confined within the semiclassical approximation.

Let us consider a massless particle in the spacetime (\ref{rind1}) described by the Klein-Gordon equation
\begin{eqnarray}
-\frac{\hbar^2}{\sqrt{-g}}\partial_\mu\left[g^{\mu\nu}\sqrt{-g}\partial_{\nu}\right]\phi = 0.
\label{1.02}
\end{eqnarray}
For radial trajectories, only the $(x-t)$ sector of the metric (\ref{rind1}) is important. Therefore under this metric the Klein-Gordon equation reduces to
\begin{eqnarray}
-\frac{1}{\Lambda x}\partial^2_t\phi +\Lambda \partial_x\phi + \Lambda x\partial_x^2\phi=0,
\label{1.03}
\end{eqnarray}
where $\Lambda$ is the constant parameter of the given Rindler metric (\ref{rind1}). The wave function satisfying the above equation is obtained by making the standard WKB ansatz for $\phi$ which is
\begin{eqnarray}
\phi(x,t)={\textrm{exp}}\left[-\frac{i}{\hbar}S(x,t)\right],
\label{1.04}
\end{eqnarray}  
where $S(x,t)$ is the single particle action.
To incorporate quantum corrections in $\phi(x,t)$ one has to expand $S(x,t)$ in powers of $\hbar$
\begin{eqnarray}
S(x,t)=S_0(x,t)+\sum_i \hbar^i S_i(x,t), &\qquad& \textrm{where} \; i=1,2,3,\ldots
\label{1.06}
\end{eqnarray}
Here our analysis will be confined only to the semiclassical approximation, i.e. $\hbar\rightarrow 0$. The effect of higher orders in $\hbar$ will be discussed in Section \ref{SecUCorr}. Substituting (\ref{1.04}) into the wave equation (\ref{1.03}), we obtain
\begin{eqnarray}
&&\frac{i}{\Lambda x}\left(\frac{\partial S}{\partial t}\right)^2 - i\Lambda x\left(\frac{\partial S}{\partial x}\right)^2 - \frac{\hbar}{\Lambda x}\frac{\partial^2 S}{\partial t^2} + \hbar \Lambda x\frac{\partial^2 S}{\partial x^2}+\hbar\Lambda\frac{\partial S}{\partial x}=0.
\label{1.05}
\end{eqnarray}
Now using (\ref{1.06}) in (\ref{1.05}) and taking the limit $\hbar\rightarrow 0$ we obtain the semiclassical Hamilton Jacobi equation
\begin{eqnarray}
\frac{\partial S_0}{\partial t}=\pm \Lambda x\frac{\partial S_0}{\partial x},
\label{1.08}
\end{eqnarray}

Since the metric (\ref{rind1}) is stationary it has timelike Killing vectors. Thus we will look for solutions of (\ref{1.08}) which behave as 
\begin{eqnarray}
S_0(x,t)=\Omega t + \tilde S_0(x),
\label{2}
\end{eqnarray} 
where $\Omega$ is a constant of motion corresponding to the timelike Killing vector. In a general curved spacetime, $\Omega$ is the product of the particle's energy $E$ as measured by an arbitrary observer and the appropriate redshift factor $V$ \cite{Carrol}. Here since $V$ equals $\sqrt{|g_{00}|}$, for the metric (\ref{rind1}), $\Omega=E  \sqrt{\Lambda x}$. We note that since the energy $E$ gets redshifted by the red-shift factor $\sqrt{|g_{00}|}$, the $x$ dependence cancels out and $\Omega$ is thus a constant of motion.

Inserting (\ref{2}) in (\ref{1.08}) and then integrating we obtain, 
\begin{eqnarray}
\tilde{S_0}(x) =  \pm \Omega\int\frac{dx}{\Lambda x}
\label{3}
\end{eqnarray} 
where the limits of the integration are chosen such that the particle goes through the horizon $x=0$ (see fig. \ref{wedgedia}). The $+ (-)$ sign in front of the integral indicates that the particle is ingoing (outgoing). Therefore substituting (\ref{3}) in (\ref{2}) we get two solutions for $S_0(x,t)$:
\begin{eqnarray}
(S_0)_{\textrm {in}}(x,t)= \Omega t + \Omega\int \frac{dx}{\Lambda x}
\label{1.10}
\end{eqnarray}
and
\begin{eqnarray}
(S_0)_{\textrm {out}}(x,t)= \Omega t  - \Omega\int \frac{dx}{\Lambda x}.
\label{1.101}
\end{eqnarray}

Now for tunneling of a particle across the horizon, the temporal and spatial natures of the coordinates $t$ and $x$ gets interchanged. Actually the two patches across the horizon are connected by a discrete imaginary amount of time \cite{Akhmedov}. This indicates that the $t$ coordinate has an imaginary part due to crossing of the horizon and correspondingly there will be a temporal contribution to the probabilities of the ingoing and outgoing particles. On using the expressions for $(S_0)_\textrm{in}(x,t)$ and $(S_0)_\textrm{out}(x,t)$ in (\ref{1.04}) the respective probabilities can be written as
\begin{eqnarray}
P_\textrm{in}&=& \left|\phi_\textrm{in}\right|^2 = ~\left| e^{-\frac{i}{\hbar} (S_0)_{\tiny \textrm{in}}} \right|^2 =\textrm{exp}\left[\frac{2}{\hbar}~\Omega ~\textrm{Im}\left(t+\int \frac{dx}{\Lambda x}\right)\right],
\label{pin}
\\
P_\textrm{out}&=& \left|\phi_\textrm{out}\right|^2 = ~\left| e^{-\frac{i}{\hbar} (S_0)_{\tiny \textrm{out}}} \right|^2 =\textrm{exp}\left[\frac{2}{\hbar}~\Omega ~\textrm{Im}\left(t-\int \frac{dx}{\Lambda x}\right)\right].
\label{pout}
\end{eqnarray}

Since in the classical limit (i.e. $\hbar\rightarrow 0$) the probability for the ingoing particle ($P_{\textrm{in}}$) has to be unity as was explained earlier, we obtain from(\ref{pin}), 
\begin{eqnarray}
{\textrm{Im}}~t = -{\textrm{Im}}\int \frac{dx}{\Lambda x} = -\frac{\pi}{\Lambda}
\label{1.16}
\end{eqnarray}
which is precisely the imaginary part of the transformation $t\rightarrow t-i\frac{\pi}{\Lambda}$ when one connects the two regions across the horizon as shown in \cite{Akhmedov}. Therefore the probability of the outgoing particle is
\begin{eqnarray}
P_{{\textrm{out}}}={\textrm{exp}}\left[-\frac{4\pi}{\Lambda\hbar}\Omega\right].
\label{1.17}
\end{eqnarray}

Now an observer observing this tunneling process will see the same particle tunneling with energy $E$ and a temperature $T_U$. Therefore when using the principle of ``detailed balance,'' \cite{Paddy2,Bibhas} the observer gets
\begin{eqnarray}
P_{{\textrm{out}}}={\textrm{exp}}\left(-\frac{E}{T_U}\right)P_{\textrm{in}}.
\label{1.18}
\end{eqnarray}
Now substituting for $P_\textrm{out}$ from (\ref{1.17}) and noting that $P_\textrm{in}=1$ we have
\begin{eqnarray}
\textrm{exp}\left[-\frac{E}{T_U}\right]=\textrm{exp}\left[-\frac{4\pi}{\Lambda\hbar}E\sqrt{\Lambda x}\right],
\label{compare}
\end{eqnarray}
where $\Omega=E~\sqrt{\Lambda x}$ has been used.
The  temperature $T_U$ is precisely the Unruh temperature
\begin{eqnarray}
T_U &=&\frac{\Lambda\hbar}{4\pi} \frac{1}{\sqrt{\Lambda x}}.
\label{1.19}
\end{eqnarray}
In terms of the local acceleration $\alpha$ given by (\ref{ourparam}), we can rewrite the Unruh temperature as
\begin{eqnarray}
T_U &=& \frac{\alpha\hbar}{2\pi}.
\label{TUn}
\end{eqnarray}
The result in this form, but obtained from completely different considerations, was given by Unruh \cite{Unruh}. In relation to the Tolman condition, which states that the product of temperature and redshift factor remains constant at all spacetime points, we see that our result is consistent. Here the redshift factor is $\sqrt{|g_{00}|}=\sqrt{\Lambda x}$ and so from (\ref{TUn}) we have $T_U~\sqrt{\Lambda x}=\frac{\Lambda \hbar}{4 \pi}$, a constant. Also note that the temperature (\ref{TUn}) seen by a Rindler observer near the Rindler horizon ($x=0$) is divergent. This is consistent with the fact that no finite energy emission from the horizon reaches beyond, as can be easily seen from redshift considerations \cite{Carrol}. A static observer at some other spacetime point, will actually see a finite, red-shifted version of this temperature as the redshift factor $\sqrt{|g_{00}|}$ appropriately cancels out the $\frac{1}{\sqrt{x}}$ term responsible for divergence. This is discussed in details in Section 5.

\label{SecU}

\section{Corrected Unruh effect by quantum tunneling}

In the previous section, the analysis was confined only at the semiclassical level reproducing the known expression (\ref{TUn}). The present discussion will go beyond the semiclassical approximation \cite{Majhibeyond, Majhiback, Majhifermi, Modak}. For this, all orders in $\hbar$ in (\ref{1.06}) and (\ref{1.05}) will be considered. Substituting (\ref{1.06}) in (\ref{1.05}) we obtain,
\begin{eqnarray}
&&\frac{i}{\Lambda x}\left(\sum_{n=0}^{\infty}\hbar^n\frac{\partial S_n}{\partial t}\right)\left(\sum_{m=0}^{\infty}\hbar^m\frac{\partial S_m}{\partial t}\right)-i\Lambda x\left(\sum_{n=0}^{\infty}\hbar^n\frac{\partial S_n}{\partial x}\right)\left(\sum_{m=0}^{\infty}\hbar^m\frac{\partial S_m}{\partial x}\right)
\nonumber
\\
&&-\frac{\hbar}{\Lambda x}\sum_{n=0}^{\infty}\hbar^n\frac{\partial^2 S_n}{\partial t^2}+\hbar \Lambda x\sum_{n=0}^{\infty}\hbar^n\frac{\partial^2 S_n}{\partial x^2}+\hbar \Lambda \sum_{n=0}^{\infty}\hbar^n\frac{\partial S_n}{\partial x}=0
\label{new1} 
\end{eqnarray}
Equating $\hbar^0$, $\hbar^1$ and $\hbar^2$ coefficients on both sides of the above equation we get the following three relations,
\begin{eqnarray}
\hbar^0:&& \left(\frac{\partial S_0}{\partial t}\right)^2 - (\Lambda x)^2\left(\frac{\partial S_0}{\partial x}\right)^2=0
\nonumber
\\
\hbar^1:&& 2i \frac{\partial S_0}{\partial t}\frac{\partial S_1}{\partial t}-2i (\Lambda x)^2 \frac{\partial S_0}{\partial x} \frac{\partial S_1}{\partial x}-\frac{\partial^2 S_0}{\partial t^2}+\Lambda^2x\frac{\partial S_0}{\partial x}+(\Lambda x)^2\frac{\partial^2 S_0}{\partial x^2}=0
\nonumber
\\
\hbar^2:&& 2i \frac{\partial S_0}{\partial t}\frac{\partial S_2}{\partial t}-2i (\Lambda x)^2 \frac{\partial S_0}{\partial x} \frac{\partial S_2}{\partial x}+i\left(\frac{\partial S_1}{\partial t}\right)^2-\frac{\partial^2 S_1}{\partial t^2}
\nonumber
\\&+&\Lambda^2x\frac{\partial S_1}{\partial x}+(\Lambda x)^2\frac{\partial^2 S_1}{\partial x^2}-i(\Lambda x)^2\left(\frac{\partial S_1}{\partial x}\right)^2=0
\label{new2}
\end{eqnarray}
Now any equation in the above set can be simplified by using the equations coming before it. This leads to the following identical set of equations:
\begin{eqnarray}
\hbar^0:&& \frac{\partial S_0}{\partial t}=\pm\Lambda x\frac{\partial S_0}{\partial x}
\nonumber
\\
\hbar^1:&& \frac{\partial S_1}{\partial t}=\pm\Lambda x\frac{\partial S_1}{\partial x}
\nonumber
\\
\hbar^2: && \frac{\partial S_2}{\partial t}=\pm\Lambda x\frac{\partial S_2}{\partial x}
\label{new31}
\end{eqnarray}
Next using the method of induction it will be shown that the above behaviour holds for any arbitrary order of $\hbar$.  For this we assume that this holds upto $\hbar^{r-1}$ order, i.e.
\begin{eqnarray}
\frac{\partial S_{r-1}}{\partial t}=\pm\Lambda x\frac{\partial S_{r-1}}{\partial x}
\label{new3}
\end{eqnarray}
where $r=1,2,3,\ldots, r$. Equating the coefficient of arbitrary $\hbar^{r}$ order on both sides of equation (\ref{new1}) we obtain,
\begin{eqnarray}
\hbar^r: i\sum_{n=0}^{r}\left(\frac{\partial S_n}{\partial t}\frac{\partial S_{r-n}}{\partial t}\right)-i(\Lambda x)^2\sum_{n=0}^{r}\left(\frac{\partial S_n}{\partial x}\frac{\partial S_{r-n}}{\partial x}\right)-\frac{\partial^2S_{r-1}}{\partial t^2}+\Lambda^2 x\frac{\partial S_{r-1}}{\partial x}+(\Lambda x)^2\frac{\partial^2 S_{r-1}}{\partial x^2}=0
\label{new4}
\end{eqnarray}
Use of (\ref{new3}) in the above immediately yields,
\begin{eqnarray}
\sum_{n=0}^{r}\left(\frac{\partial S_n}{\partial t}\frac{\partial S_{r-n}}{\partial t}\right)=(\Lambda x)^2\sum_{n=0}^{r}\left(\frac{\partial S_n}{\partial x}\frac{\partial S_{r-n}}{\partial x}\right)
\label{new5}
\end{eqnarray}
which again can be rewritten as,
\begin{eqnarray}
\sum_{n=1}^{r-1}\left(\frac{\partial S_n}{\partial t}\frac{\partial S_{r-n}}{\partial t}\right)+2\frac{\partial S_0}{\partial t}\frac{\partial S_{r}}{\partial t}=(\Lambda x)^2\sum_{n=1}^{r-1}\left(\frac{\partial S_n}{\partial x}\frac{\partial S_{r-n}}{\partial x}\right)+2(\Lambda x)^2\frac{\partial S_0}{\partial x}\frac{\partial S_{r}}{\partial x}
\label{new6}
\end{eqnarray}
Using (\ref{new3}) in the first term on left hand side leads to the cherished form,
\begin{eqnarray}
\frac{\partial S_r}{\partial t}=\pm\Lambda x\frac{\partial S_r}{\partial x}
\label{new7}
\end{eqnarray}
The whole analysis reveals that if this identical behaviour holds upto order $\hbar^{r-1}$, then it is also true for order $\hbar^r$. Therefore one can immediately conclude that the functional form of all equations in any order of $\hbar$ is identical, i.e.
\begin{eqnarray}
\frac{\partial S_a}{\partial t}=\pm\Lambda x\frac{\partial S_a}{\partial x}
\label{new8}
\end{eqnarray}
where $a=0,1,2,3,\cdots$. Consequently the solutions of these equations are not independent and $S_i$'s are proportional to $S_0$. Since $S_0$ has the dimension of $\hbar$ the proportionality constants should have the dimension of inverse of $\hbar^i$. Again in the units $G=c=k_B=1$ the Planck constant $\hbar$ is of the order of square of the Planck length $l_P$. Specifically, for this type of metric (\ref{rind1}) having $\Lambda$ (which has dimension of inverse length (see (\ref{ourparam})), as the only macroscopic parameter, these considerations show that the most general expression for $S$, following from (\ref{1.06}), is given by,
\begin{eqnarray}
S(x,t)=\left(1+\sum_iC_i\hbar^i\Lambda^{2i}\right)~S_0(x,t).
\label{new9}
\end{eqnarray}
where $C_i$'s are, as yet, undetermined dimensionless constant parameters.

To obtain a solution for $S(x,t)$ it is therefore enough to solve for $S_0(x,t)$ which satisfies $a=0$ equation of (\ref{new8}). In fact the standard Hamilton-Jacobi solution determined by this $S_0(x,t)$ is just modified by a prefactor to yield the complete solution for $S(x,t)$. The solution for $S_0(x,t)$ has been done in the previous section. Using the solutions (\ref{1.10}), (\ref{1.101}) and following identical steps employed earlier, we obtain the corrected form of the Unruh temperature:
\begin{eqnarray}
T_{U}^{(c)}=\frac{\hbar\alpha}{2\pi}\left(1+\sum_i C_i\hbar^i\Lambda^{2i}\right)^{-1}
\label{new10}
\end{eqnarray}
i.e. the semiclassical Unruh temperature is just modified by a multiplicative factor. For $C_i=0$ this reduces to the semiclassical value (\ref{TUn}) which was deduced earlier in the previous section. The parameter $\Lambda$ is obtained from the given Rindler metric. For example, for the specific Rindler metric (\ref{unruhRind}) used by Unruh in \cite{Unruh}, $\Lambda=1$ and we get the corrected Unruh temperature as
\begin{eqnarray}
T_{U}^{(c)}=\frac{\hbar\alpha}{2\pi}\left(1+\sum_i C_i\hbar^i\right)^{-1},
\label{new10inUnruh}
\end{eqnarray}
where the acceleration $\alpha$ for this metric (\ref{unruhRind}) is now $\frac{1}{2 \sqrt{x}}$.

\label{SecUCorr}

\section{Connection between corrected Unruh and corrected Hawking temperature}

In this section we show the relation between the Unruh and Hawking effects  through two different methods and exploit this relation to fix the undetermined proportionality constant of the corrected Unruh temperature. We also make  a comparative study of the two methods used. The content is divided into three subsections. In the first subsection we use the standard method of near horizon expansion of a black hole followed by redshift \cite{Unruh, Carrol} and in the second, we use the method of GEMS (Global Embedding Minkowski Spacetime) \cite{Rosen, Deser, Fronsdal, Song}. Finally in the last subsection we show how both the methods, GEMS and the standard method of relating Unruh and Hawking temperatures, are equivalent.

\subsection{Method of near horizon approximation}

In the first approach we show how a black hole metric reduces to Rindler form in a near horizon expansion. So the static observer here becomes a Rindler observer, accelerating w.r.t. the freely falling locally Minkowskian frame. Thus from the analysis of Section \ref{SecUCorr}, we see that the Unruh effect occurs and using (\ref{new10}) we get the corresponding corrected temperature. Then by redshifting this corrected Unruh temperature from horizon to infinity, the corrected form of the Hawking temperature is reproduced (to $\hbar$ order) and thereby the value of the undetermined coefficient $C_1$ is fixed for a specific black hole metric (Schwarzschild metric).  We start with a general static, spherically symmetric metric
\begin{eqnarray}
ds^2 = -f(r)dt^2 + \frac{dr^2}{f(r)} + r^2d\Omega^2,
\label{genbh}
\end{eqnarray}
whose horizon is located at $r=r_h$ where $f(r_h)=0$. If we use the Taylor expansion of the metric function about the horizon $f(r_h + \delta r) \simeq f'(r_h)\delta r + \mathcal{O}(\delta r^2)$, we get the near horizon approximation to the metric as
\begin{eqnarray}
ds^2 = -f'(r_h)x dt^2 + \frac{dx^2}{f'(r_h)x} + (r_h+x)^2 d\Omega^2,
\label{nearrind}
\end{eqnarray}
where we have made a transformation of the radial coordinate $x=r-r_h$ with the restriction $x\ll r_h$.

Comparing this with (\ref{rind1}), we see that (\ref{nearrind}) is in the Rindler form with acceleration $\alpha$, proper time $\tau$ and parameter $\Lambda$ (\ref{accn} \& \ref{ourparam}) are given by,
\begin{eqnarray}
\alpha &=& \frac{1}{F(x)} \; = \; \frac{\sqrt{f'(r_h)}}{2\sqrt x}\nonumber\\
\tau &=& \frac{[f'(r_h)/2]}{\alpha} ~t\nonumber\\
\Lambda &=& f'(r_h).
\label{nearaccn}
\end{eqnarray}
The horizon in this metric (\ref{nearrind}) corresponds to $x=0$ or $r=r_h$, i.e. the Rindler accelerated horizon ($x=0$) maps to the Schwarzschild horizon ($r=r_h$).

Thus the corrected Unruh temperature for this Rindler metric (\ref{nearrind}) with the acceleration given as in (\ref{nearaccn}), upto one-loop corrections ($C_i=0,\;\forall ~i>1$), is read off from (\ref{new10}),
\begin{eqnarray}
T^{(c)}_U &=& \frac{\hbar\sqrt{f'(r_h)}}{4\pi\sqrt x} \left( 1-C_1\hbar \left[f'(r_h)\right]^2\right)
\label{genTcor}
\end{eqnarray}
where we have used a binomial expansion. This is the temperature seen by a  static observer near the black hole horizon and is a manifestation of the Unruh Effect. The corresponding Minkowski observer here would be the freely-falling observer, who according to the principle of equivalence would observe the spacetime near horizon as locally $\left( x\ll r_h \right)$ Minkowski flat.

Now a static observer at infinity will see a redshifted Unruh temperature which is nothing but the Hawking temperature $T_H$
\begin{eqnarray}
T_H = T_{\infty} = \frac{\lim_{x \to 0} \left(V_x T_x \right)}{V_{\infty}},
\label{genred}
\end{eqnarray}
where V is the redshift factor given by $V=\sqrt{|g_{00}|}$ (see \cite{Carrol}).
Here $V_x$ is calculated from the near horizon approximated Rindler metric (\ref{nearrind}) and $V_\infty$ from the black hole metric (\ref{genbh}). Substituting appropriately we get
\begin{eqnarray}
T^{(c)}_H &=& \frac{\hbar f'(r_h)}{4\pi} \left( 1-C_1\hbar \left[f'(r_h)\right]^2\right),
\label{corred}
\end{eqnarray}
where we have imposed the asymptotic flatness property on the black hole metric.
For the standard Schwarzschild black hole of mass $M$, the metric
\begin{eqnarray}
ds^2=-\left(1-\frac{2M}{r}\right)dt^2 + \frac{dr^2}{\left(1-\frac{2M}{r}\right)} + r^2d\Omega^2
\label{schwarz}
\end{eqnarray}
gives $f'(r_h)=1/2M$ and so the corrected Hawking temperature turns out to be
\begin{eqnarray}
T^{(c)}_H &=& \frac{\hbar}{8\pi M} \left( 1-\frac{C_1\hbar}{4M^2} \right),
\label{schhawk}
\end{eqnarray}
On removing the one loop correction effect by setting $C_1=0$ we get back the standard expression for Hawking temperature as is evident from (\ref{schhawk}). Now the one-loop corrected Hawking temperature has been discussed in the literature. For example, for a Schwarzschild black hole the corrected Hawking temperature found by other methods \cite{Fursaev, Lousto, Majhibeyond} is
\begin{eqnarray}
T_H^{(c)}=\frac{\hbar}{8\pi M} \left(1-\frac{\beta_1\hbar}{M^2}\right),
\label{hawkcorr}
\end{eqnarray}
where $\beta_1$ is a constant related to trace anomaly \cite{Fursaev, Lousto, Majhiback}. The general form of $\beta_1$ is given by \cite{Fursaev},
\begin{eqnarray}
\beta_1 &=& -\frac{1}{360\pi}\left(-~N_0-\frac{7}{4}~N_{1/2}+13~N_1+\frac{233}{4}~N_{3/2}-212~N_2\right)
\label{fursbeta}
\end{eqnarray}
where `$N_s$' denotes the number of fields with spin `$s$'. In this case $N_0=1$ and $N_{1/2}=N_1=N_{3/2}=N_2=0$ giving
\begin{eqnarray}
\beta_1=\frac{1}{360\pi}.
\label{beta1}
\end{eqnarray}
Comparing (\ref{schhawk}) and (\ref{hawkcorr}) we then get (for a Schwarzschild black hole),
\begin{eqnarray}
C_1 = 4\beta_1 = \frac{1}{90\pi}.
\label{c1}
\end{eqnarray}
So the one-loop corrected Hawking temperature is obtained from (\ref{schhawk}) as
\begin{eqnarray}
T^{(c)}_H &=& \frac{\hbar}{8\pi M} \left( 1-\frac{\hbar}{360\pi M^2}\right)
\label{hawkcorr2}
\end{eqnarray}
Finally, the one-loop corrected Unruh temperature for the Rindler metric (\ref{nearrind}) arising in a near horizon expansion of the Schwarzschild metric is obtained by using the value of $C_1$ (\ref{c1}) in (\ref{genTcor})
\begin{eqnarray}
T^{(c)}_U &=& \frac{\hbar\alpha}{2\pi} \left( 1-\frac{\hbar \left[f'(r_h)\right]^2}{90\pi}\right),
\label{TUcorr1st}
\end{eqnarray}
where $\alpha$ is the local acceleration defined in (\ref{nearaccn}).

\subsection{Method of Global Embedding Minkowski Spacetime - GEMS}

Another interesting method of mapping the Unruh and Hawking temperatures is through finding a suitable embedding of the black hole spacetime in an appropriate (pseudo)Euclidean space of some higher dimension $d$. It is then demonstrated that in this higher dimensional flat metric, the black hole observers become accelerated Rindler observers, thereby leading to the Unruh effect. Now the Hawking temperature can be found corresponding to the observer at infinity either through red-shift or even directly (as we also show) if the embedding transformations are well defined at infinity. This approach has the name GEMS - Global Embedding Minkowski Spacetime, and such an embedding can always be found \cite{Goenner}. The flat space embedding for a Schwarzschild black hole is done in a 6 dimensional space \cite{Fronsdal}
\begin{eqnarray}
ds^2 &=& -\left(dz^0\right)^2+\left(dz^1\right)^2+\left(dz^2\right)^2+\left(dz^3\right)^2+\left(dz^4\right)^2+\left(dz^5\right)^2
\label{GEMSmetric}
\end{eqnarray}
The transformations (which can be extended to cover the region $r<2M$, \textit{Eq. (16) of} \cite{Fronsdal}) leading to the embedding of the Schwarzschild metric (\ref{schwarz}) in this space are
\begin{eqnarray}
z^0 &=& 4M\sqrt{1-2M/r}~\sinh(t/4M),\nonumber\\
z^1 &=& 4M\sqrt{1-2M/r}~\cosh(t/4M),\nonumber\\
z^2 &=& \int dr\sqrt{(2Mr^2+4M^2r+8M^3)/r^3},\nonumber\\
z^3 &=& r\sin \theta \sin \phi ,\nonumber\\
z^4 &=& r \sin \theta \cos \phi ,\nonumber\\
z^5 &=& r \cos \theta.
\label{GEMStrans}
\end{eqnarray}
This transformation ensures that on substitution in the GEMS flat metric (\ref{GEMSmetric}) we get back the 4 dimensional Schwarzschild metric. Observers in Schwarzschild spacetime located at some spatial point with fixed spatial coordinates $(r, \theta, \phi)$, here represent observers having fixed $z^2, z^3, z^4, z^5$ coordinates with the other two coordinates related as 
\begin{eqnarray}
\left(z^1\right)^2-\left(z^0\right)^2 &=& 16M^2\left(1-\frac{2M}{r}\right)\quad=\quad \frac{1}{\alpha_6^2}.
\label{GEMShyper}
\end{eqnarray}
On comparison with the hyperbolic trajectories (\ref{hyper}) of an accelerated observer in Minkowski space, we see that (\ref{GEMShyper}) represent similar trajectories in the $z^0-z^1$ plane. Thus different observers located at different values of $r$ represent, in the $z^0-z^1$ plane, different hyperbolae. This fact and the form of the first two transformations in (\ref{GEMStrans}) indicates that $(t,r)$ behave as Rindler coordinates modelling an accelerated observer in $z^0-z^1$ plane. This is verified by explicitly transforming the metric of the $z^0-z^1$ sector into the corresponding metric in the coordinates $(r,t)$
\begin{eqnarray}
ds^2=-\left(1-\frac{2M}{r}\right)~dt^2+\frac{16 M^4}{r^4\left(1-\frac{2M}{r}\right)}~dr^2.
\label{GEMSrtrind}
\end{eqnarray}
This metric is in Rindler form as can be seen using our generalised Rindler metric (\ref{genrind}) which readily gives the above metric (\ref{GEMSrtrind}) on choosing the function $F$ as $F(r)=4M~\sqrt{1-\frac{2M}{r}}$ and the constant $a=\frac{1}{4M}$. However, since the Rindler metric (\ref{GEMSrtrind}) is not in the form of the metric (\ref{rind1}) which was used in our tunneling analysis ({\it Section \ref{SecU} \& \ref{SecUCorr}}), we choose to introduce a new Rindler coordinate pair $(p,q)$ defined as
\begin{eqnarray}
z^0 &=& \frac{2\sqrt{q}}{\sqrt{\Lambda}}\sinh\left(\frac{\Lambda p}{2}\right)\nonumber\\
z^1 &=& \frac{2\sqrt{q}}{\sqrt{\Lambda}}\cosh\left(\frac{\Lambda p}{2}\right).
\label{GEMSrtrans}
\end{eqnarray}
The metric in the $z^0-z^1$ sector of (\ref{GEMSmetric}) is now expressed as
\begin{eqnarray}
ds^2 &=& -(\Lambda q)~dp^2+\frac{dq^2}{(\Lambda q)}
\label{GEMSrind}
\end{eqnarray}
which is in the exact Rindler form as (\ref{rind1}). On comparison with the first two relations of (\ref{GEMStrans}) with the transformations (\ref{GEMSrtrans}), we see that the coordinates $(p,q)$ and $(t,r)$ are related as
\begin{eqnarray}
p&=&\frac{t}{2\Lambda M}\nonumber\\
q&=&4M^2\Lambda\left(1-\frac{2M}{r}\right);\qquad\qquad \Lambda = \frac{1}{2M}.
\label{GEMSandSch}
\end{eqnarray}
If we substitute for coordinates ($p,q$) using the above transformation (\ref{GEMSandSch}), in the metric (\ref{GEMSrind}), we get back the metric (\ref{GEMSrtrind}) which is given in terms of ($t,r$).

The whole analysis done previously to compute the corrected Unruh temperature now goes through using the metric (\ref{GEMSrind}), and we get the corrected Unruh temperature through GEMS as below
\begin{eqnarray}
T_U^{(c)}\Big|_{GEMS}=\frac{\hbar\alpha_6}{2\pi}\left(1+\sum_i C_i\hbar^i\Lambda^{2i}\right)^{-1}.
\label{GEMStuc}
\end{eqnarray}
Upto first order in $\hbar$, this gives
\begin{eqnarray}
T_U^{(c)}\Big|_{GEMS}=\frac{\hbar}{8\pi M}\left(1-\frac{2M}{r}\right)^{-1/2}\left(1- \frac{\hbar}{360\pi M^2}\right)
\label{GEMSTuCorr1}
\end{eqnarray}
where we have substituted for $\alpha_6$ from (\ref{GEMShyper}), $C_1$ from (\ref{c1}) and $\Lambda$ from (\ref{GEMSandSch}). This is the temperature as seen by an embedded GEMS observer at an arbitrary $r$. To get the Hawking temperature from here, we again use the redshift equation (\ref{genred}) between $r$ and $\infty$ to obtain the corrected form,
\begin{eqnarray}
T_H^{(c)} &=& T_U^{(c)} \sqrt{|g_{00}|}\nonumber\\
&=& \frac{\hbar}{8\pi M}\left(1- \frac{\hbar}{360\pi M^2}\right)
\label{GEMDhawk}
\end{eqnarray}
where we have used the $g_{00}$ component of the metric (\ref{GEMSrtrind}) as the temperature (\ref{GEMSTuCorr1}) was expressed in the coordinate $r$. This reproduces the expression (\ref{hawkcorr2}).

We note that since the embedding done through the transformation (\ref{GEMStrans}) remains well-defined all the way to $r=\infty$, we can also investigate the Hawking temperature directly from the temperature (\ref{GEMSTuCorr1}). The point is that the Unruh temperature (\ref{GEMSTuCorr1}) varies from one hyperbolae to the other in $z^0-z^1$ plane, depending upon the value of $r$ (\ref{GEMShyper}) and each hyperbola represents accelerated observers in $(z^0,z^1)$ coordinates having acceleration $\alpha_6 = \frac{1}{4M}\left(1-\frac{2M}{r}\right)^{-1/2}$. The Hawking observer is the observer who is situated at $r=\infty$. So here in this case the Unruh temperature (\ref{GEMSTuCorr1}) seen at $\infty$ is precisely equal to the Hawking temperature. This is demonstrated quantitatively when we use $r=\infty$ in the expression (\ref{GEMSTuCorr1}) to recover the corrected Hawking temperature (\ref{hawkcorr2}) or (\ref{GEMDhawk}).

\subsection{The GEMS and the near horizon expansion - a comparative study}

As we have shown, both methods -- the near horizon expansion followed by redshift, and the method of embedding the black hole in a flat spacetime (GEMS) -- provide a map between Unruh and Hawking effects. Now since both methods are being used to describe the same physical map, we can easily anticipate that the two methods must relate with one another, such that results derived using both are identical. We explore this point further in this subsection.

The corrected Unruh temperature found near the horizon in (\ref{TUcorr1st}) can be red-shifted to give $T(r)$ -- the temperature observed at an arbitrary position $r$ -- through use of the redshift equation (\ref{genred}) as
\begin{eqnarray}
T(r) &=& \frac{T_U^{(c)}(r_h)~V(r_h)}{V(r)}\nonumber\\
&=& \left[\frac{\hbar\sqrt{f'(r_h)}}{4\pi\sqrt x} \left( 1-\frac{\hbar \left[f'(r_h)\right]^2}{90\pi}\right)~\sqrt{f'(r_h) x}~\right]\Big/ \sqrt{f(r)},
\end{eqnarray}
where we have substituted for the redshift factor $V=\sqrt{|g_{00}|}$ from the appropriate metrics near the respective points under consideration. Simplifying this and taking the case of the Schwarzschild metric, where $f(r)=\left(1-\frac{2M}{r}\right)$ and $f'(r_h)=\frac{1}{2M}$, we get
\begin{eqnarray}
T(r) &=& \frac{\hbar}{8\pi M}~\left(1- \frac{\hbar}{360\pi M^2}\right)~\left(1-\frac{2M}{r}\right)^{-1/2},
\label{Unruhatr}
\end{eqnarray}
which is the exact expression as found through an embedding analysis (GEMS) in (\ref{GEMSTuCorr1}). Now we can redshift this to $\infty$ by $T(r)\sqrt{g_{00}}$  to recover our expression for the corrected Hawking temperature (\ref{GEMDhawk}) or (\ref{hawkcorr2}). So we see that both the methods give the same temperature as observed at any intermediate point $r$ between the horizon and infinity. This demonstrates the inter-compatibility of the two methods.

\section{Conclusion and discussions}

In this paper, we make a systematic study of the Unruh effect in the tunneling formalism through a Hamilton-Jacobi analysis. Since there is a close connection between Unruh and Hawking effects, and the latter has been studied extensively in the tunneling formalism, it is natural to consider this approach for the Unruh effect also. Apart from illuminating several points in the connection between these two effects, we also found that the Unruh effect undergoes corrections, exactly like the Hawking effect. This is a new result.

We did a Hamilton-Jacobi tunneling analysis within the WKB ansatz (but taking higher orders of $\hbar$ to obtain quantum corrections) to describe a massless scalar particle tunneling through the Rindler accelerated horizon ({\it as shown in fig.~\ref{wedgedia}}). While doing so, our analysis shows how a careful choice of the form of the single particle action $S_0$ (\ref{2}) and application of the principle of detailed balance in conjunction with the redshift relation gave the observer dependant Unruh temperature. Confining to the leading WKB term ($\hbar \rightarrow 0$) we obtained the familiar result of the Unruh temperature as $T=\frac{\alpha}{2\pi}$, where $\alpha$ is the local observer's acceleration. We are thus able to reproduce the original result of Unruh \cite{Unruh} in the tunneling picture.

We next concentrated on the corrections to the Unruh temperature and found a corrected form (\ref{new10}). Also, by mapping the corrected Unruh temperature to the corrected Hawking temperature through redshift arguments, we were able to fix the coefficient to the leading order correction term.

Furthermore, our analysis clarified some aspects of the tunneling picture in a black hole, when we studied the connection between the Unruh and Hawking temperatures. The Rindler horizon was shown to map to the Schwarzschild event horizon and the picture of scalar-particle tunneling developed here, successfully maps to that of the well known case of tunneling in black holes. The Unruh temperature previously found thus applies to a near horizon approximated Schwarzschild metric which has been shown to take a Rindler form. Redshift arguments then gave the temperature $T(r)$ at any arbitrary coordinate distance $r$ (see (\ref{GEMSTuCorr1}) or (\ref{Unruhatr}) ). This temperature corresponds to a `local Hawking temperature' in the nomenclature used by Deser and Levin \cite{Deser}. The standard Hawking temperature is obtained by looking at the asymptotic limit $r\rightarrow \infty$.

Besides the method of connecting the Unruh and Hawking effects through a near horizon approximation, there exists another approach (GEMS) where the Schwarzschild metric is treated as a 4-D hypersurface of a higher, 6-D Minkowski-flat space. Different detectors at different distances $r$ in the black hole metric get mapped into different accelerated observer hyperbolae in the higher dimensional coordinate plane ($z^0-z^1$ plane in our example). Thus we have the Unruh effect, and the Rindler observer corresponding to $r \rightarrow \infty$ sees the Hawking temperature. The embedding however has to be such that the hypersurface should contain the black hole event-horizon, as stressed in \cite{Deser}. Without this, the temperature vanishes \cite{Deser}. Here we note how the horizon also becomes important in the tunneling picture. The horizon there appears as a singularity which gives a pole in the integral (\ref{1.16}) (or more generally in (\ref{genImtime})) without which there will be no imaginary temporal contribution and no temperature. In a comparative study between the two methods, we saw how both methods gave the same corrected Unruh temperature observed at arbitrary $r$, and from there, the Hawking temperature observed at $\infty$ was found through redshift for both. Thus we have clearly demonstrated the equivalence of the two methods. Either one can be used, though we note that a difficult part in the embedding method is to find a \emph{globally valid} embedding transformation in the first place. So it seems easier to follow the method of near horizon approximation.

At the end, we note that the tunneling calculations were done in a specific form of the Rindler metric. This was desirable because the near horizon approximation of the Schwarzschild black hole resulted in that particular Rindler metric. However, as pointed in Appendix B, it is feasible to perform this analysis starting from the generalised Rindler metric (\ref{genrind}) that we had constructed earlier.

\begin{appendix}

\renewcommand{\thesection}{Appendix \Alph{section} :}			% redefine the command that creates the section heading.
\setcounter{section}{0}											% redefine the command that creates the section no.

\section{The generalised Rindler metric applied in some standard cases}

\renewcommand{\theequation}{A.\arabic{equation}} % redefine the command that creates the equation no.
\setcounter{equation}{0}  % reset counter 

The Generalised Rindler metric (\ref{genrind})
\begin{eqnarray}
ds^2 = -a^2 F(x)^2 dt^2 + F'(x)^2 dx^2 + dy^2 + dz^2.\nonumber
\end{eqnarray}
yields various forms of the Rindler metric through specific choice of the function $F(x)$ and the constant $a$. Some examples of standard forms of Rindler metrics found in literature are discussed below.

\paragraph*{}

{\bf Metric 1.}
\begin{equation}
ds^2 = e^{2 \alpha x} \left(-dt^2 + dx^2\right).
\label{carRind}
\end{equation}
For this metric taken from \cite{Carrol}, we get $a=\alpha$ and $F(x)=\frac{1}{a}e^{a x}$.

\paragraph*{}

{\bf Metric 2.}
\begin{equation}
ds^2 = - x^2 \, dt^2 + dx^2.
\label{rindRind}
\end{equation}
For this metric taken from \cite{Rindler}, we get $a=1$ and $F(x)=x$.

\paragraph*{}

{\bf Metric 3.}
\begin{equation}
ds^2 = - x \, dt^2 + \frac{dx^2}{x}.
\label{unruhRind}
\end{equation}
For this metric taken from \cite{Unruh}, we get $a=1/2$ and $F(x)=2\sqrt{x}$.\\
\vspace {4mm}\\
Finally, an example of a metric that is not in Rindler form:

\paragraph*{}

{\bf Metric 4.}
\begin{equation}
ds^2 = - x^2 \, dt^2 + \frac{dx^2}{x^2}.
\label{timbuctoo}
\end{equation}
For this metric, no choice of $F(x)$ and a {\it constant} `$a$' exists that allow us to write it in the form of (\ref{genrind}). Thus this is not a valid Rindler metric.

For all metrics described in this appendix except (\ref{timbuctoo}), the corresponding Minkowski Space variables are defined through the transformations (\ref{rtrns}).

\section{Tunneling in the generalised Rindler metric}

\renewcommand{\theequation}{B.\arabic{equation}} % redefine the command that creates the equation no.
\setcounter{equation}{0}  % reset counter 

In Section \ref{SecU} we had taken an explicit Rindler metric (\ref{rind1}) keeping in mind that our near horizon approximated metric (\ref{nearrind}) comes naturally in that form. There it had lead to a simple, transparent analysis as against taking the generalised form (\ref{genrind}) which we had constructed earlier. This is because, the function $F$ in the general form, is an arbitrary function of $x$, with only general restrictions that it be analytic, monotonic and positive definite. However, the tunneling analysis can be carried out in the generalised form itself, as we now outline briefly, stating only the relevant results. We consider the standard semiclassical ($\hbar \rightarrow 0$) case, from which it can be seen that an analysis along the lines of Section \ref{SecUCorr}, with all powers of $\hbar$ corrections included, is feasible.

The Klein Gordon equation (\ref{1.02}) for our massless scalar particle, in the generalised metric (\ref{genrind}), considering the relevant $t-x$ sector, turns out to be
\begin{eqnarray}
-\frac{1}{a^2F^2}\left(\partial_t^2 \phi\right)+\frac{1}{F'^2}\left(\partial_x^2\phi\right)+\frac{1}{F F'} \left(\partial_x\phi\right)-\frac{F''}{F'^3}\left(\partial_x\phi\right)=0
\label{genKG}
\end{eqnarray}
The standard WKB ansatz (\ref{1.04}) for the wave function $\phi$, when inserted in (\ref{genKG}) yields, on taking the $\hbar \rightarrow 0$ limit,
\begin{eqnarray}
\left(\frac{\partial S}{\partial t}\right)=\pm \frac{aF}{F'} ~\left(\frac{\partial S}{\partial x}\right)
\label{genCherish}
\end{eqnarray}
where $S\equiv S(x,t)$ is the single particle action.
Now taking a similar decomposition of the action as in (\ref{2}) and inserting in equation (\ref{genCherish}) yields the ingoing (+) and outgoing (-) modes $\phi_\textrm{in}$ and $\phi_\textrm{out}$:
\begin{eqnarray}
\phi_\textrm{in} &=& e^{-\frac{i}{\hbar}S_{\tiny \textrm{in}}}\;=\; \mathrm{exp}\left[-\frac{i}{\hbar}\left(\Omega t+\frac{\Omega}{a}\int \frac{dF}{F}\right)\right]\nonumber\\
\phi_\textrm{out} &=& e^{-\frac{i}{\hbar}S_{\tiny \textrm{out}}}\;=\; \mathrm{exp}\left[-\frac{i}{\hbar}\left(\Omega t-\frac{\Omega}{a}\int \frac{dF}{F}\right)\right]\nonumber\\
\label{genphimode}
\end{eqnarray}
The ingoing and outgoing probabilities are calculated as $|\phi|^2$, taking the relevant modes. Again imposing the condition $P_\textrm{in}=1$ as explained earlier, we get $\textrm{Im}~ S_{\footnotesize \textrm{in}}=0$, leading to the result,
\begin{eqnarray}
\textrm{Im}~ t = - \textrm{Im}~ \frac{1}{a} \int \frac{dF}{F} = -\frac{\pi}{a}
\label{genImtime}
\end{eqnarray}
where the limits of the integral were taken just across the zero of $F(x)$ (say at $x=x_0$). The zero of the function $F(x)$ gives us the position of the horizon. Subsequent analysis of using this result (\ref{genImtime}) to calculate the outgoing probability and then using the principle of detailed balance now goes through as before. Finally we get the familiar form of the Unruh temperature
\begin{eqnarray}
T_U=\frac{\alpha\hbar}{2\pi},
\label{genTu}
\end{eqnarray}
where $\alpha$ is here the local acceleration, and is related to the function $F$ as $\alpha=\frac{1}{F}$ ({\it see} \ref{accn}).

\end{appendix}

\end{document}